\documentclass[lettersize,journal]{IEEEtran}
\usepackage{amsmath,amsfonts}
\usepackage{algorithmic}
\usepackage{algorithm}
\usepackage{array}
\usepackage[caption=false,font=normalsize,labelfont=sf,textfont=sf]{subfig}
\usepackage{textcomp}
\usepackage{stfloats}
\usepackage{url}
\usepackage{verbatim}
\usepackage{graphicx}
\usepackage{cite}
\usepackage{hyperref}

\usepackage{multirow}
\usepackage{bm}
\usepackage{makecell}
\usepackage{adjustbox}

\usepackage{pifont}
\usepackage{color}
\usepackage[normalem]{ulem}
\useunder{\uline}{\ul}{}
\begin{document}

\title{A Scanning Prior Guided Super-Resolution Framework for Photoacoustic Microscopy Images with Comprehensive Evaluations}

\author{Kai Pan, Linyang Li, Li Lin, Pujin Cheng, Junyan Lyu, Lei Xi, and Xiaoying Tang, \IEEEmembership{Senior Member, IEEE}
\thanks{This study was supported by the National Key Research and Development Program of China (2023YFC2415400); the National Natural Science Foundation of China (62071210, 62022037, 61775028, 81571722, 61528401); the Guangdong Basic and Applied Basic Research (2024B1515020088); the Department of Science and Technology of Guangdong Province (2019ZT08Y191, 2022B1212010003); the Shenzhen Science and Technology Program (RCYX20210609103056042, RCJC20231211090039066, 20231116104616001, KQTD20190929172743294);  the Shenzhen Science and Technology Innovation Committee (KCXFZ2020122117340001); the Guangdong Basic and Applied Basic Research (2021A1515220131); the High Level of Special Funds (G030230001, G03034K003); the Startup grant from Southern University of Science and Technology (PDJH2021C008). \emph{(Corresponding authors: Lei Xi, xilei@sustech.edu.cn; Xiaoying Tang, tangxy@sustech.edu.cn.)}}
\thanks{Kai Pan, Li Lin, Pujin Cheng, Junyan Lyu, and Xiaoyin Tang are with the Department of Electronic and Electrical Engineering, Southern University of Science and Technology, Shenzhen, China. Li Lin, Pujin Cheng, and Xiaoying Tang are also with Jiaxing Research Institute, Southern University of Science and Technology, Jiaxing, China}
\thanks{Linyang Li and Lei Xi are with the Department of Biomedical Engineering, Southern University of Science and Technology, Shenzhen, China.}
\thanks{Kai Pan and Linyang Li contributed equally to this work.}}

\markboth{IEEE TRANSACTIONS ON INSTRUMENTATION AND MEASUREMENT}%
{Shell \MakeLowercase{\textit{et al.}}: A Sample Article Using IEEEtran.cls for IEEE Journals}


\maketitle

\begin{abstract}
Photoacoustic microscopy (PAM) has emerged as a valuable tool in clinical practice, leveraging the combined advantages of optical contrast and acoustic resolution. However, a critical challenge lies in balancing imaging speed with resolution, hindering its practical application. Recently, deep learning methods have shown promise in accelerating PAM imaging via super-resolution techniques. Yet, in most studies, the evaluation of the generated images' quality mainly relies on pixel-wise metrics such as the Peak Signal-to-Noise Ratio based on synthetic datasets, with almost none evaluation on downstream tasks nor real datasets. Moreover, the acquisition of PAM images involves various scanning schemes, including rotational and raster scanning, resulting in diverse imaging modes. Existing methods often neglect prior knowledge on the scanning scheme, failing to incorporate such prior into their models. In this study, we propose a novel super-resolution framework for PAM images integrating scanning prior. Our framework addresses several key challenges. First, we introduce an efficient registration module to align image displacements caused by scanner wobbling. Additionally, we propose a gradient-based patch selection strategy to prioritize blood vessel patches during training, enhancing our model's ability to capture important features. Furthermore, we introduce a scanning consistency loss to embed scanning-specific characteristics into the training process. Finally, a downstream segmentation task is employed to comprehensively evaluate the quality of the generated images. Experimental results on both synthetic and real datasets demonstrate the effectiveness and generalizability of our proposed framework for PAM image super-resolution. Code is available at \href{https://github.com/11710615/PAMSR.git}{https://github.com/11710615/PAMSR.git}.
\end{abstract}


\section{Introduction}\label{sec:introduction}
Photoacoustic microscopy (PAM) imaging is an emerging optical imaging technique and has been widely applied in functional brain imaging analyses \cite{pam12, pam13, pam11}. Compared to other imaging techniques, such as functional magnetic resonance imaging \cite{pam3, pam4} and optical microscopy \cite{pam7,pam5, pam6}, PAM imaging provides a high resolution, a large field of view (FOV), and a deep penetration depth for real-time imaging of the hemodynamics in intricate brain vascular networks \cite{pam8,pam9, pam10}. 

High-resolution (HR) PAM images providing clear vessel structures are crucial for \textit{in vivo} animal studies and clinical applications. However, the commonly-employed point-scanning scheme in PAM imaging often requires a large number of scanning points and a small scanning step size to obtain HR images, leading to prolonged imaging time. Additionally, the imaging speed is usually limited by the laser's pulse repetition rate to ensure compliance with the safety thresholds of optical energy density. As a result, sparse sampling has become a necessary compromise to accelerate imaging.

\begin{figure}[htbp]
\centering
\includegraphics[width=0.95\columnwidth]{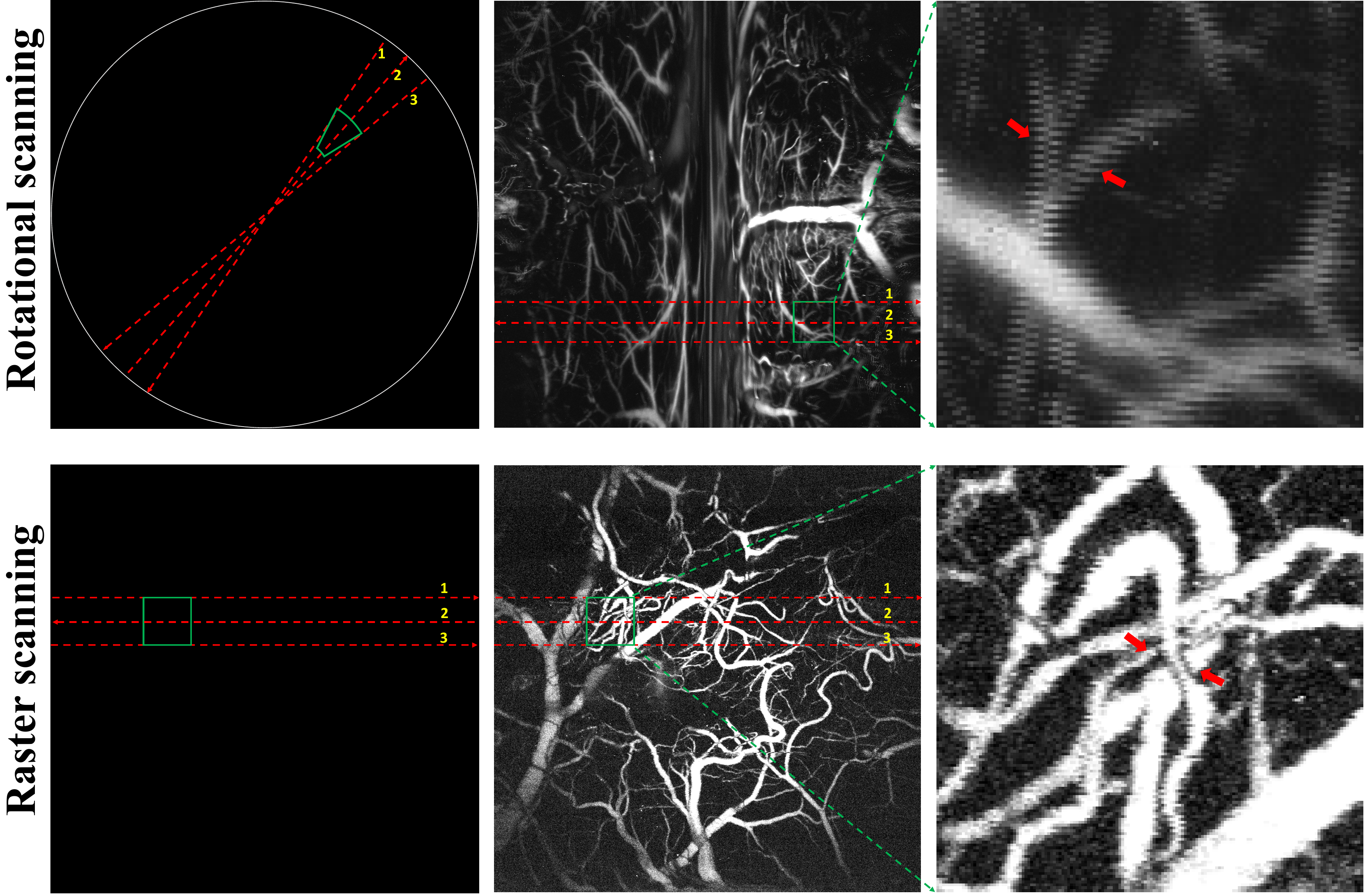}
\caption{Inherent displacement degradation existed in PAM images. Both rotational scanning and raster scanning encounter displacements between adjacent rows.}
\label{1a}
\end{figure}

PAM imaging employs various schemes to fulfill different application needs, including rotational \cite{pam22,pam23, pam24} and raster scanning \cite{pam11, cao2023optical}. In rotational scanning, the scanner moves along diameters, while in raster scanning, it follows a grid pattern, as illustrated in Fig. \ref{1a}. Those two scanning schemes employ different sparse sampling strategies. For example, raster scanning achieves undersampling by increasing the scanning step size, whereas rotational scanning simultaneously increases both step size and angle intervals. 

Recently, deep learning (DL)-based super-resolution (SR) methods have made significant advancements in the natural image analysis field \cite{zamir2022restormer, liang2021swinir, edsr, rdn, dat}, offering potential for accelerating PAM imaging \cite{zhou2021photoacoustic, dispirito2020reconstructing, vu2021deep}.
By applying these SR methods to the undersampled image obtained with a large scanning step size, the quality of the SR image can approach that of full sampling, thereby accelerating the imaging process. However, several challenges hinder the direct transfer of SR methods to PAM images. Firstly, existing SR methods are primarily designed for natural images and may not be able to effectively address the specific degradations inherent in PAM imaging. PAM images inevitably suffer from displacements within the image of interest due to scanning imperfections. For instance, high-speed scanning can lead to position offsets between adjacent scanning lines, resulting in displacements across rows \cite{zhu2022real, gao2022achieving, hong2024unsupervised}, as shown in Fig. \ref{1a}. Hence, aligning these displacements before applying SR methods becomes necessary. Secondly, the acquisition of PAM images is more challenging than that of natural images, resulting in smaller training datasets. Therefore, designing modules or strategies for efficient training becomes imperative. Thirdly, PAM images play functional roles in clinical studies. Existing evaluation metrics such as the Peak Signal-to-Noise Ratio (PSNR) primarily assess image quality at the pixel level, which may not sufficiently capture the nuances that are critical for clinical evaluations. Consequently, it is of great necessity to develop novel and effective PAM-specific SR pipelines along with comprehensive evaluation analyses. 

In this study, we introduce a scanning prior guided SR framework for accelerating PAM imaging. Our framework encompasses registration-highlighted preprocessing, patch-selection and scanning prior guided training, and downstream oriented evaluation. Specifically, an efficient registration module is devised in the preprocessing stage to align displacements caused by scanner wobbling. During training, a gradient-based patch selection strategy is used to prioritize blood vessel patches, and a scanning consistency loss is employed to integrate scanning-specific prior knowledge. Finally, vessel segmentation is designed as a downstream task for more comprehensive evaluation. Experimental results on both synthetic and real undersampled PAM datasets demonstrate the superiority and robustness of our proposed framework.


\section{Related Work}
\subsection{Single-image Super-resolution}
\subsubsection{CNN-based Methods}
A single-image SR (SISR) task focuses on recovering HR images from their low-resolution (LR) counterparts. In recent years, the development of convolutional neural networks (CNNs) has significantly advanced SISR methods.  For instance, Ledig \emph{et al.} \cite{srresnet} propose SRResNet, which uses residual blocks to recover fine texture details. Lim \emph{et al.} \cite{edsr} remove batch normalization layers from residual blocks to prevent high-frequency components' recovery from being hindered. Zhang \emph{et al.} \cite{rdn} design a residual dense block to fully utilize hierarchical features. Zhang \emph{et al.} \cite{rcan} propose the RCAN model, which combines the channel attention mechanism with residual blocks to adaptively weigh features. Niu \emph{et al.} \cite{han} introduce the holistic attention network, which includes a layer attention module and a channel-spatial attention module to rescale feature weights for informative representations. Dai \emph{et al.} \cite{san} propose a non-locally enhanced residual group to expand the receptive field and emphasize contextual information. Mei \emph{et al.} \cite{paedsr} propose a pyramid attention module that fully utilizes self-similarity across multi-scale features. 

In addition to various residual and attention module designs, edge information has also been incorporated into SISR methods. For instance, Zhu \emph{et al.} \cite{zhu2015modeling} utilize the composition and deformation of LR gradient patterns to estimate HR gradient fields for more reliable SR results. Yang \emph{et al.} \cite{yang2017deep} propose an edge-based SR approach that embeds prior edge information into their network. Ma \emph{et al.} \cite{ma2020structure} design a gradient reconstruction branch to guide the SR model's training, achieving excellent results in preserving structural similarity. Liang \emph{et al.} \cite{liang2022details} employ residual variance to identify artifacts in the restored images and reduce the weight of the artifact regions in its model's loss function, which strategy can be easily incorporated into existing methods to enhance performance.

\subsubsection{Transformer-based Methods}
The aforementioned CNN methods suffer from limited receptive fields. Recently, transformer-based frameworks have emerged as state-of-the-art approaches in image reconstruction tasks. Compared with CNNs, transformers perform better in establishing long-range dependencies and have larger receptive fields. Liang \emph{et al.} \cite{liang2021swinir} apply the Swin-Transformer model \cite{liu2021swin} to the SR, denoising, JPEG compression, and artifact removing tasks, achieving impressive performance. Wang \emph{et al.} \cite{wang2022uformer} propose a U-shaped transformer structure with a non-overlapping window-based self-attention mechanism to reduce computation and a multi-scale feature adjustment modulator. Zamir1 \emph{et al.} \cite{zamir2022restormer} propose a Restormer model focusing on multi-scale local-global representation learning, achieving great performance on several image restoration tasks. Chen \emph{et al.} \cite{hat} propose a hybrid attention transformer to activate more pixels for image super-resolution. Chen \emph{et al.} \cite{dat} apply spatial and channel self-attention in transformer blocks for more powerful feature representation.

Due to the inherent displacements within PAM images caused by scanning imperfection, it is not optimal to directly apply existing SR methods to PAM images. PAM images' inherent misalignment will confuse the training, leading the model to believe that the misalignment is an image quality issue that needs to be restored. Relatively few works take into account this unique imaging characteristic of PAM images. In this work, we propose an effective registration module to align the displacements before the SR process.

\subsection{Undersampled PAM Image Reconstruction}
The task of PAM image reconstruction is to enhance the quality of PAM images from undersampled ones. Traditional interpolation algorithms such as Bicubic and Bilinear interpolation estimate the amplitudes of unsampled points based on their neighbors, which tend to induce distortions. Alternatively, DL-based methods have shown strong representation learning capabilities for undersampled PAM images' reconstruction. For example, the Fully Dense U-net is applied to reconstruct undersampled PAM images of mouse brains' vasculature at different downsampled rates \cite{dispirito2020reconstructing}. Zhou \emph{et al.} \cite{zhou2021photoacoustic} use a residual network for optical-resolution PAM (OR-PAM) images' SR. Seong \emph{et al.} \cite{seong2023three} for the first time propose a modified SRResNet architecture to reconstruct 3D PAM data. Different from the aforementioned methods that train on paired fully-sampled and synthetic LR images, Vu \emph{et al.} \cite{vu2021deep} make use of the characteristic that recovery of the noise is slower than that of the content and HR PAM images are obtained by stopping training at a suitable iteration step. Considering the scarcity of training samples, Ma \emph{et al.} \cite{ma2024dove} utilize hand-drawn doodles to simulate blood vessel structures and employ a diffusion model to synthesize a large amount of diverse data for SR training.

Existing works mentioned above primarily focus on reconstructing undersampled PAM images without strategies to accelerate training nor modules to incorporate scanning-specific prior. In terms of evaluating the quality of the generated images, most methods employ pixel-wise similarity metrics such as the PSNR, lacking downstream tasks for comprehensive evaluation. Furthermore, most methods evaluate their frameworks solely on synthetic undersampled datasets, lacking results on real undersampled datasets, which diminishes persuasiveness for practical application. In this work, we propose a gradient-based patch selection strategy to prioritize blood vessel patches and a scanning consistency loss to integrate scanning-specific prior. Meanwhile, vessel segmentation is adopted as a downstream task for more comprehensive
evaluation. Further, we demonstrate the superiority and robustness of our proposed framework on both synthetic and real undersampled PAM datasets.

\begin{figure*} [htbp]
\centering
\includegraphics[width=0.9\textwidth]{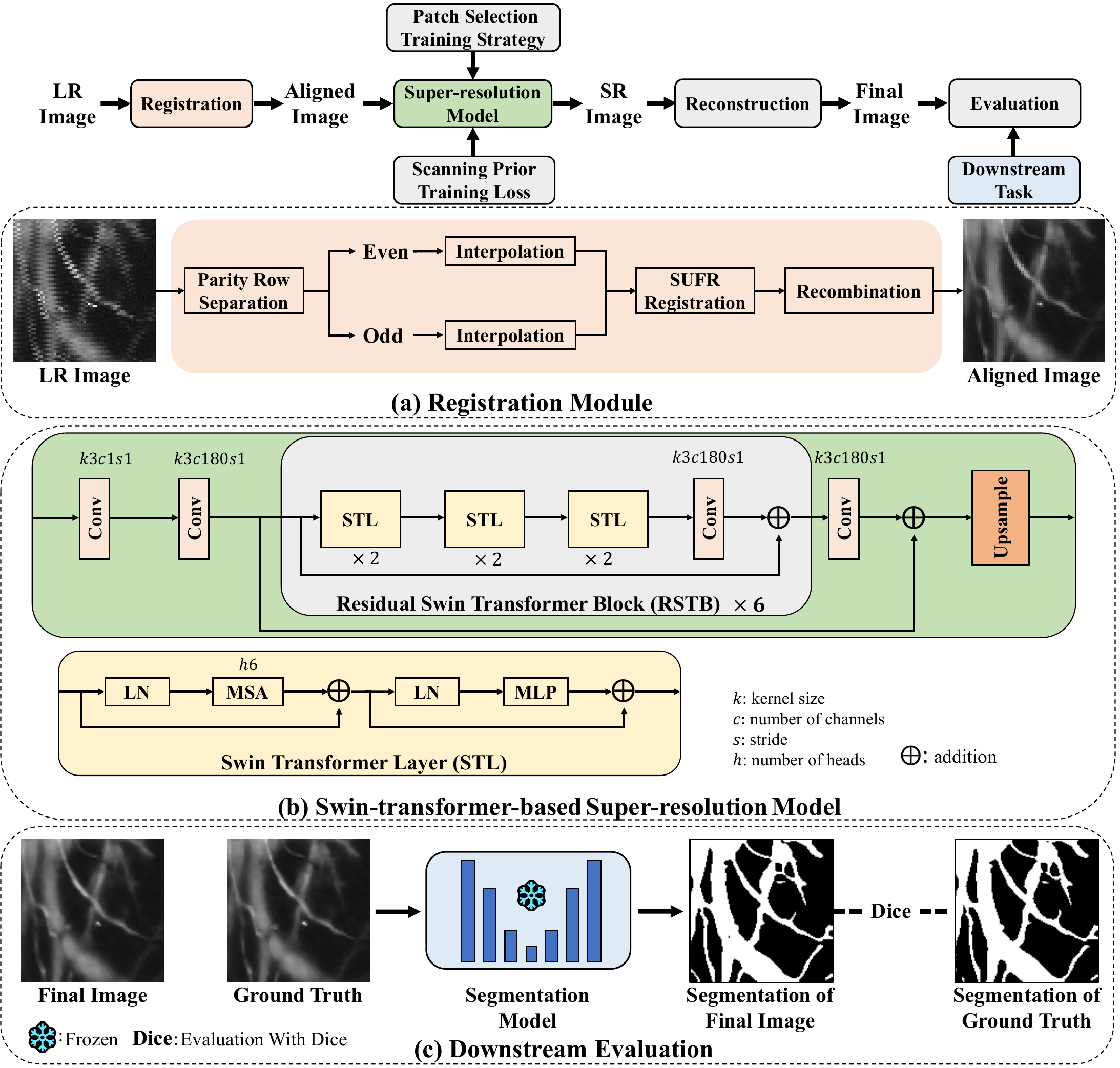}
\caption{Our proposed framework for PAM image super-resolution. (a) illustrates the registration module. (b) depicts the architecture of our training model. (c) showcases the downstream segmentation task.} 
\label{fig2}
\end{figure*}

\section{Method}
The proposed framework is illustrated in Fig. \ref{fig2}. It consists of (1) preprocessing involving image registration, (2) training with a gradient-based patch selection strategy and a scanning consistency loss, and (3) evaluation with a downstream task. Let $I_{lr}^{unreg}$, $I_{lr}$, $I_{sr}$, $I_{hr}$, and $I_{sr}^{rec}$ be the original LR image without any preprocessing, the aligned LR image, the model's output image, the corresponding HR image, and the final reconstructed image, respectively. Specifically, our objective is to train an SR model that maps $I_{lr}$ to $I_{hr}$ in the sampling space, where $I_{lr}$ is obtained from $I_{lr}^{unreg}$ processed with a proposed registration module. Distinguishing between sampling space and image space is essential for effective training and evaluation. In the sampling space, image coordinates directly correspond to the scanning process. After generating $I_{sr}$ in the sampling space, it is transformed into the image space to produce $I_{sr}^{rec}$, where pixel values are aligned to their true spatial locations through a polar-to-cartesian transformation, as shown in Fig. \ref{zero_mask}. The quality of $I_{sr}^{rec}$ is subsequently evaluated through a downstream task. In the training process, a proposed gradient-based patch selection strategy is used to prioritize blood vessel patches for enhancing the model’s ability to capture important features. Meanwhile, a consistency loss is employed to integrate scanning-specific prior knowledge into the training process.

In the following subsections, we first introduce the registration module used to align displacements within the original LR image. Then, we detail the architecture of our model. Afterwards, we introduce the gradient-based patch selection strategy and the scanning prior related consistency loss. Finally, we present the downstream task for a comprehensive evaluation purpose.

\subsection{Image Registration}
Before applying the SR model, registration is essential to correct displacements and ensure image continuity in the sampling space. Displacements often occur due to scanner wobbling, leading to misalignments between adjacent scanning lines (image rows) in the sampling space.

The registration process involves three key steps, as illustrated in Fig. \ref{fig2}(a). First, parity row separation is performed. Odd and even rows are separately sampled and linearly interpolated to match the original image shape. The interpolated odd result serves as the fixed part, while the even one represents the unregistered part.

The second step entails registration, where the unregistered part is aligned with the fixed part. This alignment is achieved using Speeded Up Robust Feature (SURF), a feature-based registration method. SURF extracts scale and rotation invariant features using a Hessian-matrix-based blob detector, facilitating the alignment accuracy.

Finally, recombination is conducted to replace the interpolated part in the odd image with the corresponding part in the registered even image. This recombination step helps minimize changes or distortions introduced by the registration process, ensuring integrity of the recombined image.

The registration process plays a crucial role in preparing PAM images for SR reconstruction, correcting displacements and ensuring the accuracy of subsequent analyses. However, whether it is a necessary step depends on whether the dataset of interest is synthetic or real. In the synthetic case, no registration is applied to the LR images because they are artificially downsampled from the registered HR images. In contrast, for the real case, registration is applied to both the HR images and its corresponding real LR images to ensure the accuracy of subsequent analyses. The aforementioned process is represented by the following equations
\begin{gather}
    \{I_{hr}, I_{lr}\}_{syn} = \{R(I_{hr}^{unreg}), D(R(I_{hr}^{unreg}))\}, \\
    \{I_{hr}, I_{lr}\}_{real} = \{R(I_{hr}^{unreg}), R(I_{lr}^{unreg})\},
\end{gather}
where $\{\cdot\}_{syn}$ and $\{\cdot\}_{real}$ respectively represent the registered synthetic and real HR-LR pairs. $R$ denotes the registration process, and $D$ represents the downsampling operation.

\subsection{Network Architecture}
The proposed network architecture is based on a transformer-based model inspired by SwinIR \cite{liang2021swinir}, aiming at capturing both local and global information from PAM images.

The network consists of three main components: shallow feature extraction, deep feature extraction, and image reconstruction, as shown in Fig. \ref{fig2}. Given an input LR image $I_{lr}$ of size $H\times W$, the two convolutional layers in the shallow feature extraction module extract local shallow features $F_{0}$ of size $H\times W\times C$, where $C$ represents the number of channels.

To capture more global and deeper features, a deep feature extraction module is then applied to the shallow features. This module is composed of a series of Residual Swin Transformer Blocks (RSTBs) and convolutional layers. The RSTBs are designed to capture long-range dependencies and global information by incorporating self-attention mechanisms. The convolutional layers further refine the features and enhance their representation power.

Finally, a reconstruction module up-samples the extracted features and generates the corresponding SR image $I_{sr}$. Here we employ PixelShuffle to directly upsample images. The whole network is formulated as:
\begin{gather}
    F_{0} = Conv(I_{lr}), \\
    F_{i} = RSTB(F_{i-1})+F_{i-1}, \    \  i=1,2,...,n,\\
    F_{deep} = Conv(F_{n})+F_{0}, \\
    I_{sr} = Rec(F_{deep}),
\end{gather}
where $Conv(\cdot)$ is a convolutional layer; $F_{i}(i \ge 1)$ is the $i$th RSTB module's output while $F_{0}$ is the shallow feature; $n$ is the total number of RSTB modules; $Rec(\cdot)$ is the reconstruction module.

RSTB contains an even number of cascaded Swin Transformer Layers (STLs), which tails with a convolutional layer in the end. STL calculates multi-head self-attention in a window and employs a shifted window mechanism for window-level feature fusion. In detail, the shallow feature $F_{shallow}$ is split into non-overlapped windows each with a size of $\frac{HW}{hw} \times hw \times C $, where $h$ and $w$ are the window's height and width. Self-attention (SA) is calculated across pixels within each window as below:
\begin{equation}
    Attention(Q,K,V)=softmax(QK^{T}/\sqrt{d}+B)V,
\end{equation}
where $Q, K, V \in \mathbb{R}^{n^2\times d}$ are the $query$, $key$, and $value$ matrices projected from the window, and $B$ is the learnable position embedding matrix. Window Multi-head Self-attention (WMSA) is performed through parallel attention layers weighed with a parameter matrix. Although WMSA saves lots of computation, there is no connection across windows as attention only represents pixel-wise correlations within the limited-size window. Therefore, Shifted-Window Multi-head Self-attention (SW-MSA) operates window shifting before conducting WMSA. For the aim of strengthening the connection across different windows, WMSA and SWMSA are applied in two continuous STLs, formulated as:
\begin{gather}
    X=WMAS(LN(X))+X,\\
    X=MLP(LN(X))+X,\\
    X=SWMAS(LN(X))+X,\\
    X=MLP(LN(X))+X,
\end{gather}
where $LN$ represents the layer-norm layer and $MLP$ represents the multi-layer perceptron layer.

\subsection{Gradient-based Patch Selection Strategy}
The proposed gradient-based patch selection strategy plays a crucial role in efficiently training the SR network by prioritizing patches containing more important information, such as regions with rich blood vessels.

In detail, the LR image $I_{lr}$ is initially divided into non-overlapping sub-windows with a size of $64\times64$. Then, the gradient information, strongly associated with vessel richness, is utilized to assign a score to each sub-window. The gradient score is calculated by applying the gradient operator to each sub-window in both the $x$ and $y$ directions. The gradient operator emphasizes the edge region in the image of interest. The score for each sub-window $X_{sub}$ is obtained by summing the gradient scores in both the $x$ and $y$ directions. A higher score indicates higher priority of the sub-window to be selected, which is calculated as follows:
\begin{equation}
    S(X_{sub}) = \sum (\frac{1}{2}G_{x}(X_{sub}) + \frac{1}{2}G_{y}(X_{sub})),
\end{equation}
where $S(\cdot)$ presents the gradient score, $G_{x}(\cdot)$ and $G_{y}(\cdot)$ are the gradient operator applied respectively to $x$ and $y$ directions.

The probability of selecting each sub-window is then determined based on the proportion of the gradient scores, formulated as:
\begin{equation}
P(X_{sub}^{i}) = \frac{{S(X_{sub}^{i})}}{{{\sum_{j=0}^{N-1}} S(X_{sub}^{j})}},
\end{equation}
where $P(X_{sub}^{i})$ is the probability for the $i$th sub-window to be selected, $N$ presents the total number of sub-windows. During training, we prioritize selecting patches within sub-windows of high priority scores, ensuring efficient training. Patches with lower scores are not entirely excluded but are selected less frequently, allowing all areas to still contribute to the training process.

\begin{figure}[tbp]
\centering
\includegraphics[width=0.95\columnwidth]{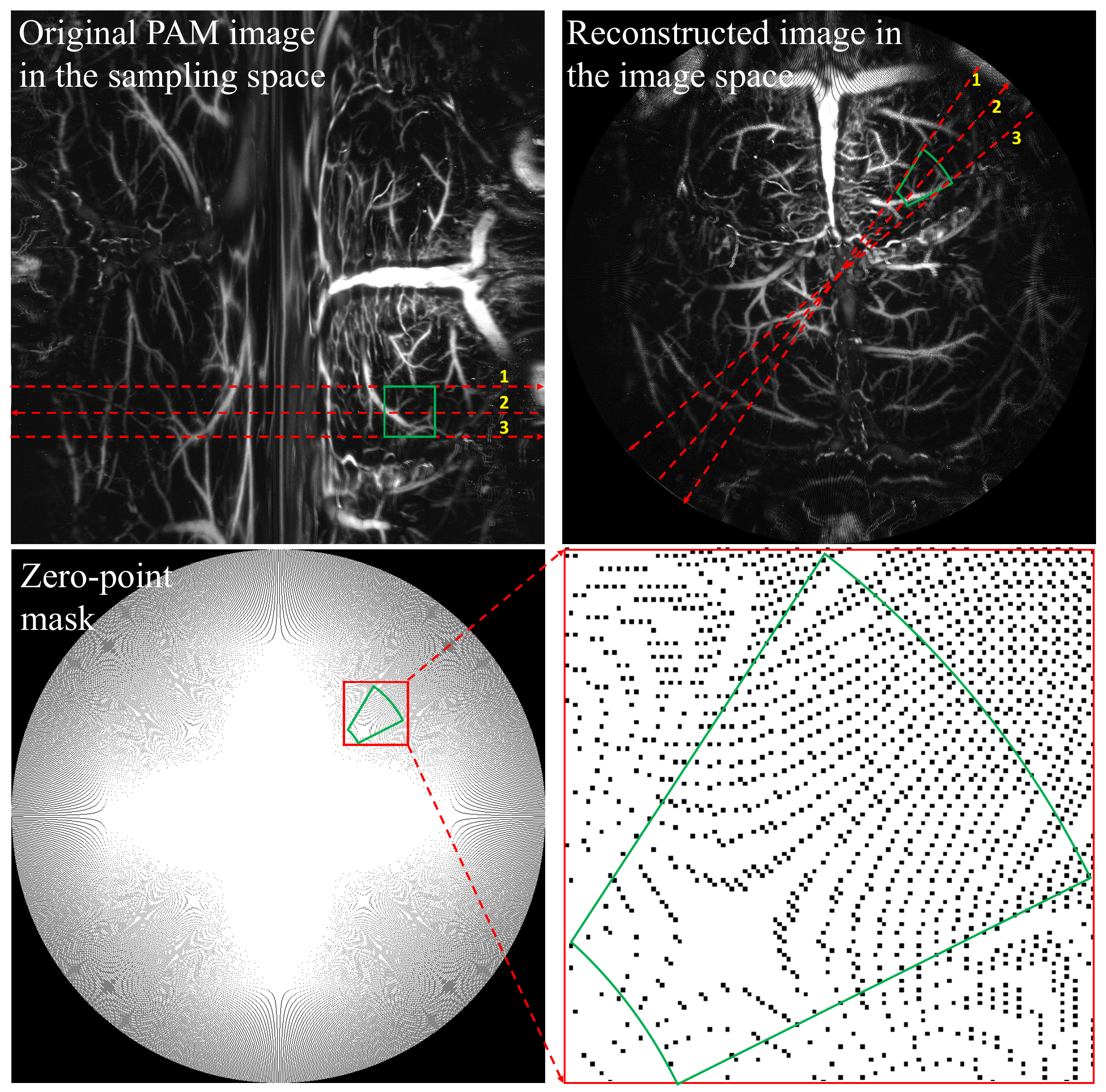}
\caption{Rotational scanning derived PAM images need to be reconstructed
from the sampling space to the image space because of the zero-point issue.}
\label{zero_mask}
\end{figure}

\subsection{Scanning Consistency Loss}
In rotational scanning, the scanner moves along diameters, sequentially collecting scanning lines to form an image in the sampling space. However, due to the inherent unevenness of this sampling process, certain regions may end up with limited or no data. Consequently, for the reconstruction process, the core lies in transforming polar coordinates to cartesian coordinates. During this transformation, the coordinate values are discretized into integers, which can cause multiple points to map to the same location due to precision limitations. This imprecision may lead to a zero-point mask, highlighting regions where valid data points are absent, as depicted in Fig. \ref{zero_mask}. The possibility of a zero-point mask makes it challenging to directly apply SR methods in the image space, as the absence of data in certain regions may degrade the quality of a generated image.

Considering this challenge, our SR model is first applied to the LR images in the sampling space, where the zero-point issue is less problematic. However, to accurately assess the quality of the generated images, it is crucial to evaluate them in the image space, where pixel values are mapped to their true locations through the reconstruction process. To improve the quality of the reconstructed images, we propose a loss $L_{sc}$ in the reconstructed image space to incorporate scanning scheme related prior knowledge into the training process. Given the importance of edge recovery in SR tasks, $L_{sc}$ is specifically designed to enforce gradient consistency between the reconstructed SR image $I_{sr}^{rec}$ and the reconstructed HR image $I_{hr}^{rec}$.

We calculate the gradients in both the $x$ and $y$ directions of an image $I$ using the gradient operator, and then assign the root mean square of both directions' gradients as the gradient map $GM(I)$. $L_{sc}$ is obtained by calculating the $L_{1}$ distance between $GM(I_{sr}^{rec})$ and $GM(I_{hr}^{rec})$ as follows:
\begin{gather}
    GM(I) = \sqrt{( G_{x}(I))^2 + (G_{y}(I))^2}, \\
    L_{sc} = E||(GM(I_{sr}^{rec}) - GM(I_{hr}^{rec})||_{1}.
\end{gather}

In addition to the scanning consistency loss $L_{sc}$, we utilize an $L_{rec}$ loss to constrain pixel-wise consistency between the SR result $I_{sr}$ and the HR image $I_{hr}$ in the sampling space,
\begin{equation}
    L_{rec} = E||I_{sr} - I_{hr}||_{1}.
\end{equation}

Therefore, the overall objective function $L_{total}$ combines the pixel-wise loss $L_{rec}$ and the scanning consistency loss $L_{sc}$ weighed by a trade-off parameter $\lambda$:
\begin{equation}
    L_{total} = L_{rec} + \lambda L_{sc}.
\end{equation}

\subsection{Evaluation Metrics}
For quantitative analysis, Peak Signal-to-Noise Ratio (PSNR), Structural Similarity Index Measure (SSIM), and Mean Square Error (MSE) are used to directly evaluate the generated PAM images' quality. PSNR and MSE measure pixel-wise distances between the SR results and the corresponding HR images, while SSIM evaluates the structural similarity. Specifically, MSE is defined as
\begin{equation}
    MSE = \frac{1}{HW}\sum_{i=0}^{H-1}\sum_{j=0}^{W-1}(I_{sr}^{rec}(i,j)-I_{hr}^{rec}(i,j))^{2},
\end{equation}
where $H$ and $W$ represent the height and width of the image, respectively. PSNR is defined as
\begin{equation}
    PSNR = 20log(\frac{I_{MAX}}{MSE}),
\end{equation}
where $I_{MAX}$ represents the maximum pixel intensity.
And, SSIM is defined as
\begin{equation}
    SSIM = \frac{{(2\mu_{I_{sr}^{rec}}\mu_{I_{hr}^{rec}} + c_1)(2\sigma_{I_{sr}^{rec}I_{hr}^{rec}} + c_2)}}{{(\mu_{I_{sr}^{rec}}^2 + \mu_{I_{hr}^{rec}}^2 + c_1)(\sigma_{I_{sr}^{rec}}^2 + \sigma_{I_{hr}^{rec}}^2 + c_2)}},
\end{equation}
where $\mu$, $\sigma$, and $\sigma_{I_{sr}^{rec}I_{hr}^{rec}}$ represent the mean, variance, and covariance, respectively. $c_1$ and $c_2$ are small positive constants utilized to stabilize computation.

PAM images play a critical role in characterizing blood vessels' morphology, such as the blood vessels' branching pattern and the vessel diameters' distribution. To assess the vessel generation performance, we conduct vessel segmentation on both the SR and HR results using a well-trained Unet model. We consider the segmentation result from the HR image as the ground truth and utilize the DICE score to evaluate the quality of the SR result, which is defined as
\begin{equation}
    DICE = \frac{2 \times |S(I_{sr}^{rec}) \cap S(I_{hr}^{rec})|}{|S(I_{sr}^{rec})| + |S(I_{hr}^{rec})|},
\end{equation}
where $S$ represents the vessel segmentation model, $|S(I_{sr}^{rec}) \cap S(I_{hr}^{rec})|$ represents the number of pixels that are correctly classified as vessels by the segmentation model in both the SR and HR images, and $|S(I_{sr}^{rec})|$ and $|S(I_{hr}^{rec})|$ represent the total numbers of vessel pixels identified by the segmentation model respectively in the SR image and the HR image.

\begin{figure*}[htbp]
\centering
\includegraphics[width=0.75\linewidth]{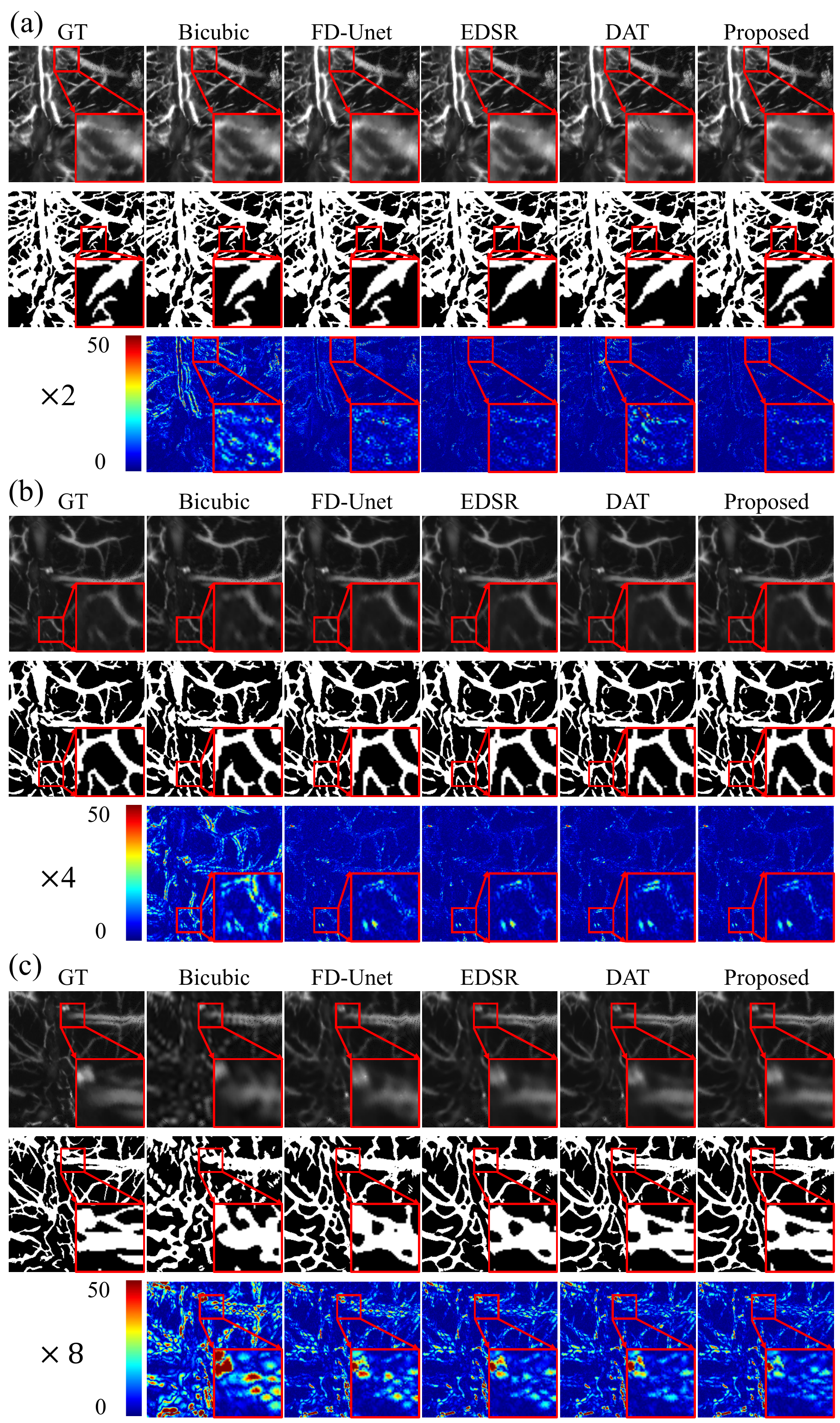}
\caption{Visual comparisons on synthetic undersampled PAM images with scale factors of 2 (a), 4 (b) and 8 (c). Each sub-figure showcases the SR results in the first row, the corresponding segmentation results in the second row, and the error maps in the third row. Regions with obvious differences are highlighted by red rectangles.}
\label{vx48_syn}
\end{figure*}

\begin{figure} [htbp]
\centering
\includegraphics[width=1\linewidth]{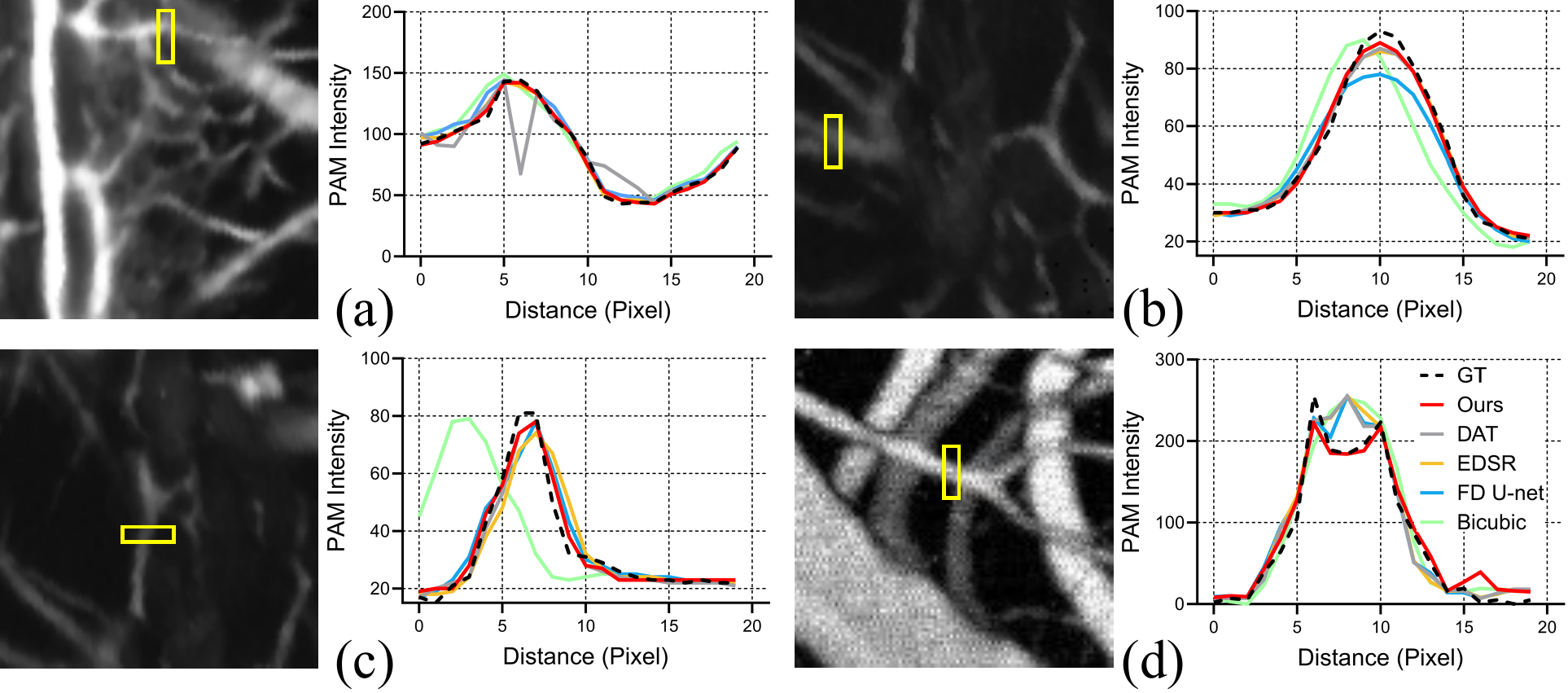}
\caption{Visual comparisons of pixel intensity profiles along three small vessels: (a) to (c) show results from synthetic data with scale factors of 2, 4, and 8, respectively, while (d) displays results from real undersampled data. Our results demonstrate the closest resemblance to the ground truth.}
\label{distance_x48}
\end{figure}


\section{Experiments}

\subsection{Datasets}
The proposed SR framework undergoes evaluations on both synthetic and real datasets, sourced from the Multi-Functional Optical Imaging Lab at Southern University of Science and Technology\footnote{The data is available at \href{https://drive.google.com/drive/folders/1yNEO2FqZFdNE9kqOV1kwfPnEifcPhH-o?usp=drive_link}{this link.}}. The synthetic dataset is composed primarily of \textit{in vivo} mouse brain microvasculature data acquired at a wavelength of $532 nm$, using the rotational scanning scheme previously published in \cite{pam12}. During each scanning process, 1000 depth-resolved PA signals (A-line) with a step interval of $10 \mu m$ in one cross-sectional image (B-scan) and 1000 B-scans with an interval angle of $0.18^{\circ}$ are collected to form a volumetric image. Then maximum amplitude projection is operated to generate a 2D PAM image with a pixel resolution of $1000 \times 1000$. For the real dataset, it is obtained using the raster scanning scheme. The LR image is obtained at a resolution of $512 \times 512$. The step size for both x and y axes is set to $6 \mu m$. The corresponding HR image comprises $1024 \times 1024$ resolution, with a step size of $3 \mu m$ along both the x-axis and the y-axis.    

The synthetic dataset consists of 116 HR images, which are randomly partitioned into 60 images for training, 16 for validation, and 40 for testing. The synthetic LR images are generated by applying nearest downsampling to the HR images. On the other hand, the real dataset comprises 41 LR-HR image pairs, intended to assess the framework's robustness in real-world scenarios. All images are saved as 8-bit grayscale images after registration and rescaled to the intensity range of [0,1] through min-max normalization for network inputs. Notably, the model is trained exclusively on the synthetic dataset and tested on both synthetic and real undersampled datasets to validate its generalizability.

Additionally, a segmentation dataset is collected to train the blood vessel segmentation model. This dataset includes 5 images featuring mouse ear blood vessel data and brain microvasculature data, acquired under conditions identical to those of the synthetic dataset. The vessel regions are manually labeled for training the segmentation model.

\subsection{Implementation Details}
In the registration process, the SURF algorithm is implemented using the RegistrationEstimator Toolbox in MATLAB, with the octave number and scale level number set to 3 and 5, respectively. Nonrigid registration is adopted with a smoothing level of 3 and a pyramid level of 1.

The SR model's architecture in our framework is built upon the SwinIR \cite{liang2021swinir}, with the RSTB number, STL number, window size, channel number and attention head number set to 6, 6, 8, 180 and 6, respectively. During the training process, the model is optimized using the ADAM optimizer with a learning rate of $10^{-4}$, $\beta_1 = 0.9$, $\beta_2 = 0.999$, and $\epsilon = 10^{-8}$ for 200 epochs. The mini-batch size is set to 8, and each image is cropped to a size of $64\times 64$ for training. For the scale factor of 2, the model is trained in an end-to-end manner with random initialization. For the scale factors of 4 and 8, the model parameters are initialized with the model trained for the scale factor of 2, except for the upsampling module which is initialized randomly. The Sobel operator is adopted to extract gradients in the patch selection process, and the weight $\lambda$ used in the loss function is set to 0.5. All models are implemented using the PyTorch framework.

\subsection{Experiment Results}
Our proposed SR framework is compared with several state-of-the-art (SOTA) methods, including EDSR \cite{edsr}, DAT \cite{dat}, and FD-Unet \cite{dispirito2020reconstructing}. Bicubic interpolation is also included in the comparison, serving as a baseline. EDSR and DAT are SR methods for enhancing natural images, while FD-Unet is specifically designed for accelerating PAM imaging. All compared methods are trained using the hyper-parameters specified in their respective original papers.

\begin{table}[tbp]
\centering
\caption{Quantitative evaluations on the synthetic undersampled dataset. The best performance for each metric is highlighted in bold and the second best one is underlined.}
\label{tab1}
\resizebox{1\columnwidth}{!}{%
\begin{tabular}{c|c|ccccc}
\hline
Scale                       & Metric & Bicubic & FD U-net\cite{dispirito2020reconstructing} & EDSR\cite{edsr}   & DAT\cite{dat}    & Ours   \\ \hline
\multirow{4}{*}{$\times 2$} & PSNR$\uparrow$    & 37.266  & 40.912   & 42.261 & \underline{42.372} & \textbf{42.935} \\
                            & SSIM$\uparrow$    & 0.984   & \underline{0.991}    & \textbf{0.993}  & \textbf{0.993}  & \textbf{0.993}  \\
                            & MSE$\downarrow$     & 13.968  & 6.374    & 5.088  & \underline{4.990}  & \textbf{4.342}  \\
                            & DICE$\uparrow$    & 92.467   & 96.387 & 97.092
  & \underline{97.100}  & \textbf{97.178}  \\ \hline
\multirow{4}{*}{$\times 4$} & PSNR$\uparrow$    & 30.641  & 36.297   & 37.625 & \underline{37.681} & \textbf{37.862} \\
                            & SSIM$\uparrow$    & 0.942   & 0.976    & \underline{0.982}  & \underline{0.982}  & \textbf{0.983}  \\
                            & MSE$\downarrow$     & 62.674  & 17.229   & 13.120 & \underline{12.958} & \textbf{12.402} \\
                            & DICE$\uparrow$    & 80.870   & 92.064    & \underline{93.165}  & 93.137& \textbf{93.195}  \\ \hline
\multirow{4}{*}{$\times 8$} & PSNR$\uparrow$    & 25.862  & 30.489   & \underline{30.919} & 30.903 & \textbf{31.197} \\
                            & SSIM$\uparrow$    & 0.865   & 0.934    & 0.937  & \underline{0.938}  & \textbf{0.941}  \\
                            & MSE$\downarrow$     & 189.458 & 65.364   & \underline{59.034} & 59.068 & \textbf{55.494} \\
                            & DICE$\uparrow$    & 65.890   & 80.071    & 80.353  & \underline{80.921}  & \textbf{81.452}  \\ \hline
\end{tabular}%
}
\end{table}

\begin{figure}[htbp]
\centering
\includegraphics[width=1\columnwidth]{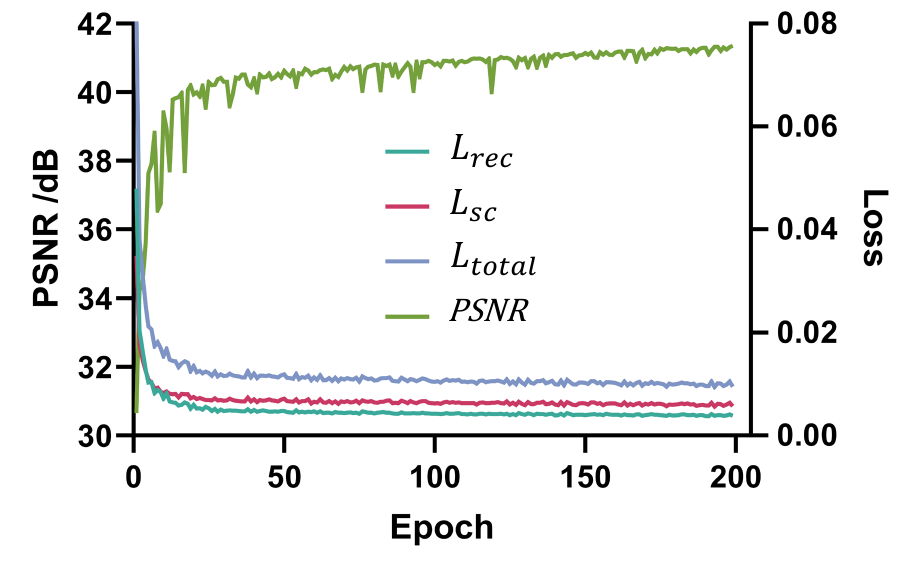}
\caption{Training loss curves and
the averaged PSNR curve on the validation set as functions of the epoch.}
\label{ablation_figure}
\end{figure}

\subsubsection{SR Results on Synthetic Undersampled Dataset}
To evaluate the SR results on the synthetic undersampled dataset, we conduct quantitative analyses at scale factors of 2, 4, and 8, as presented in Table \ref{tab1}. Our method consistently outperforms other methods across all evaluation metrics (PSNR, SSIM, MSE, and DICE) at all the three scale factors. For instance, at the 2$\times$ upscaling factor, our method achieves a PSNR of 42.935 dB, significantly surpassing other methods. While FD-Unet, EDSR, and DAT exhibit comparable SSIM scores, they notably lag behind in PSNR and MSE values compared to our method. Bicubic, on the other hand, fails to achieve comparable performance with our method at any upscaling factor. Particularly noteworthy is the improvement in the DICE score. As the upscaling factor increases, the leading advantage of our method over DAT expands from 0.078\% to 0.531\%, suggesting our method's ability to restore blood vessel details even at large upscaling factors.

Visual comparisons of the SR results are presented in Fig. \ref{vx48_syn}, with results for the scale factors of 2, 4 and 8 respectively illustrated in sub-figures (a), (b) and (c). In each sub-figure, the SR results are listed in the first row, the corresponding segmentation results are displayed in the second row, while the error maps are shown in the third row. Regions of interest with noticeable differences are highlighted with red rectangles. The error map represents the absolute value of the pixel-wise image intensity's difference between the generated result and the ground truth image, where brighter regions indicate larger deviations from the ground truth. Notably, our method exhibits the darkest blue color in the vessel edge regions, indicating minimal errors. Furthermore, our method shows the closest segmentation results to the ground truth, suggesting that the generated images closely resemble the ground truth in terms of blood vessel characteristics. Fig. \ref{distance_x48} (a), (b), and (c) show the pixel intensity profiles within the yellow boxes in the ground truth images at scale factors of 2, 4, and 8, respectively. Our method achieves the closest intensity distribution to the ground truth, indicating superior performance in recovering structural details compared to other methods.

In Fig. \ref{ablation_figure}, the proposed method's training loss curves and the averaged PSNR curve on the validation set at a scale factor of 2 are depicted as functions of the epoch. It is evident that the PSNR evolution is consistent with both the pixel-wise loss $L_{rec}$ and the consistency loss $L_{sc}$, confirming the effectiveness of our training process.

\begin{table}[tbp]
\centering
\caption{Quantitative evaluations on the real undersampled dataset. The best performance for each metric is highlighted in bold and the second best one is underlined.}
\label{real_seg}
\resizebox{1\columnwidth}{!}{%
\begin{tabular}{c|ccccc}
\hline
Metric & Bicubic  & FD U-net\cite{dispirito2020reconstructing} & EDSR\cite{edsr}   & DAT\cite{dat}      & Ours     \\ \hline
PSNR$\uparrow$    & 18.831   & \underline{19.060}   & 18.685   & 18.715   & \textbf{19.311}   \\ SSIM$\uparrow$    & 0.314    & \underline{0.325}    & 0.307    & 0.310    & \textbf{0.333}    \\
                             MSE$\downarrow$     & 1234.103 & \underline{1190.440} & 1255.837 & 1238.095 & \textbf{1113.427} \\
                             DICE$\uparrow$    & 70.156   & \underline{71.209}   & 69.905   & 69.788 & \textbf{71.461}   \\ \hline
\end{tabular}%
}
\end{table}

\subsubsection{SR Results on Real Undersampled Dataset}
\begin{figure*} [tbp]
\centering
\includegraphics[width=0.85\textwidth]{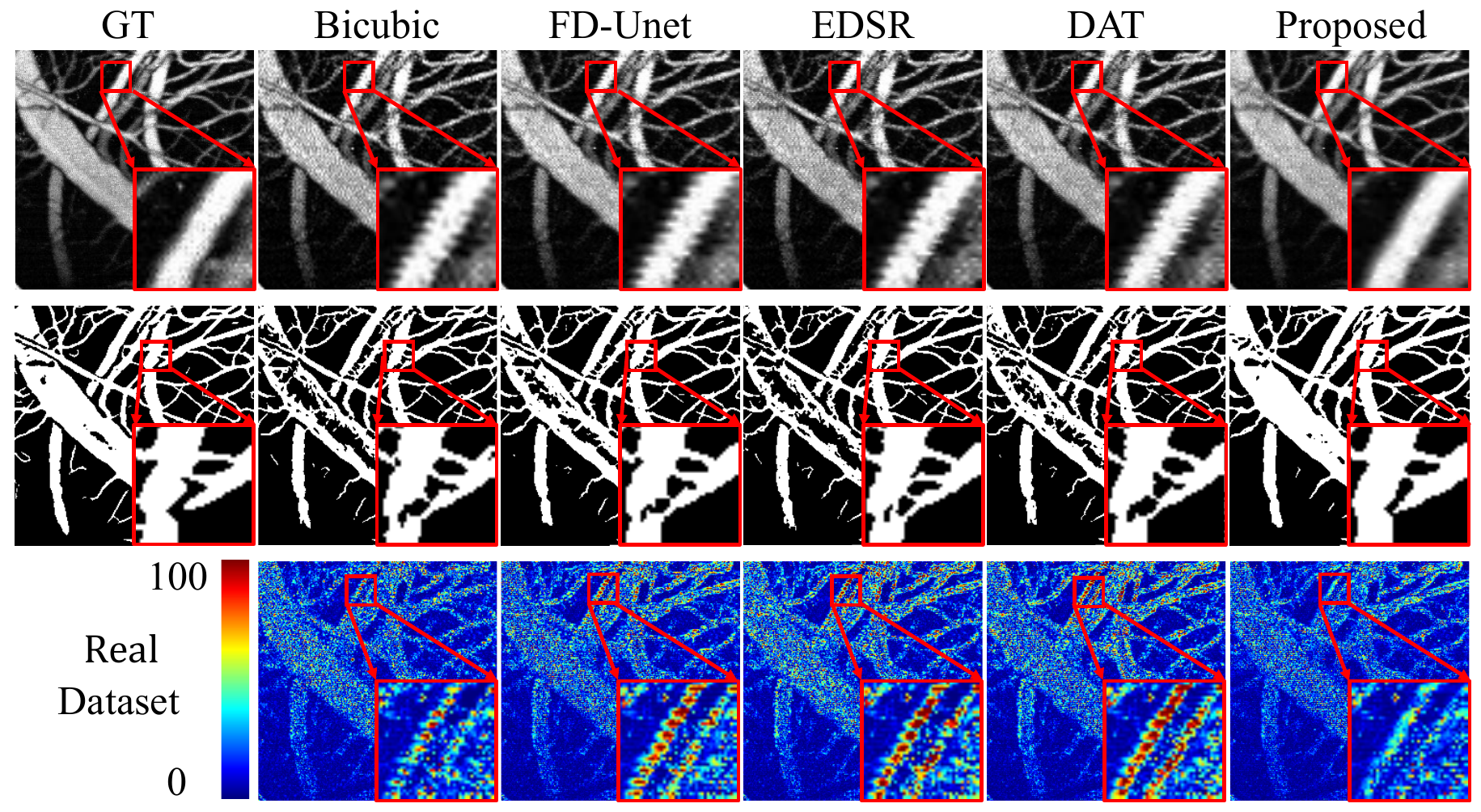}
\caption{Visual comparisons on real undersampled PAM images, showcasing SR results in the first row, corresponding segmentation results in the second row, and error maps in the third row. The regions with obvious differences are highlighted by red rectangles.}
\label{comparedreal}
\end{figure*}
We evaluate the performance of our method on real undersampled PAM images to assess its generalization capability. As described earlier, the experimental data are obtained using relatively low excitation light energy, resulting in noisy PAM images. This background noise significantly impacts traditional quantitative metrics like PSNR, making it challenging to observe clear performance differences between methods. However, downstream tasks su ch as vessel segmentation, which are less affected by background noise, can more accurately distinguish the quality of image generation. Notably, our method achieves a DICE score of 71.461\%, significantly higher than that of other methods, demonstrating its superior capability in vessel generation and robustness to noise. These results, summarized in Table \ref{real_seg}, confirm that while conventional metrics may underestimate the performance due to noise, the DICE score effectively highlights the substantial improvements made by our approach.

Visual comparisons are presented in Fig. \ref{comparedreal}, which displays the SR results in the first row, the corresponding segmentation results in the second row, and the error maps in the third row. It is evident from the error maps that our method produces smaller errors in the vessel edge regions. Additionally, the segmentation results show that our method's outputs closely approximate the HR images, particularly in the vessel branch regions. Fig. \ref{distance_x48}(c) illustrates the pixel intensity profiles within the yellow box marked in a real image. Our method is the only one that accurately captures the peak intensity, closely matching the intensity distribution in the ground truth image. 

\begin{table}[tbp]
\centering
\caption{Ablation study on the synthetic undersampled dataset with a scale factor of 2. The best performance for each metric is highlighted in bold and the second best one is underlined.}
\label{ablation}
\resizebox{1\columnwidth}{!}{%
\begin{tabular}{cccc|ccc}
\hline
\begin{tabular}[c]{@{}c@{}}Ablation\\ No.\end{tabular} & $M_{reg}$ & $M_{ps}$ & $L_{sc}$ & PSNR$\uparrow$ & MSE$\downarrow$ & DICE$\uparrow$ \\ \hline 
0& \ding{55} & \ding{55} & \ding{55} & 42.016          & 5.311          & 97.020       \\
1& \ding{51} & \ding{55} & \ding{55} & 42.642          & 4.555          & 97.056       \\
2& \ding{55} & \ding{51} & \ding{51} & 42.751          & 4.418          & 97.085       \\
3& \ding{51} & \ding{51} & \ding{55} & 42.878          & 4.374          & 97.122       \\
4& \ding{51} & \ding{55} & \ding{51} & {\ul 42.879}    & {\ul 4.368}    & {\ul 97.162} \\
Ours& \ding{51} & \ding{51} & \ding{51} & \textbf{42.935} & \textbf{4.342} & \textbf{97.178}       \\ \hline
\end{tabular}%
}
\end{table}


\subsection{Ablation Study}
To assess the importance of each module in the proposed framework, ablation studies are conducted on the synthetic undersampled dataset with a scale factor of 2. The baseline, denoted as ablation 0, represents the original SwinIR network \cite{liang2021swinir}. Ablation 1 adds the registration module $M_{reg}$ to the baseline. Ablations 2, 3, and 4 denote the proposed method without $M_{reg}$ or $L_{sc}$ or $M_{ps}$, respectively. The proposed method without $M_{reg}$ means unregistered images are used for training and the SR results are later registered for subsequent analyses.

As shown in Table \ref{ablation}, removing any individual component leads to a degradation in performance across all evaluation metrics. Specifically, when $M_{reg}$ is excluded, the proposed method shows inferior performance, resulting in decreases of 0.184 dB PSNR and 0.093\% DICE. Upon adding $M_{reg}$ to the baseline, a noticeable increase of 0.626 dB in PSNR is observed, indicating significant improvements achieved by incorporating the registration module. Additionally, the evaluation metrics consistently improve when either $M_{ps}$ or $L_{sc}$ is added to ablation 1, while they decrease when either $M_{ps}$ or $L_{sc}$ is removed. This emphasizes the significance of both $M_{ps}$ and $L_{sc}$ in enhancing the super-resolution performance of the proposed framework.


\section{Conclusion}

In this study, we introduce a comprehensive framework for accelerating PAM imaging. The framework encompasses preprocessing, training, and evaluation stages, addressing various challenges associated with PAM imaging. During the preprocessing stage, we propose a SURF-based registration module to mitigate displacements within PAM images, ensuring accuracy for subsequent analyses. In the training process, we introduce novel strategies aimed at achieving faster convergence and improved generation quality. Specifically, we propose a gradient-based patch selection strategy, prioritizing blood vessel patches to focus the model's learning on critical regions. Additionally, we introduce a scanning consistency loss to enhance the model's ability to capture fine details, further improving the overall synthesis quality. To comprehensively evaluate the generation performance, we perform a downstream vessel segmentation task, ensuring that the synthesized images are suitable for subsequent analysis and interpretation. Extensive experiments conducted on both synthetic and real undersampled datasets across different scale factors demonstrate the effectiveness and robustness of our proposed framework. Our results showcase significant improvements, underscoring the potential of our approach for accelerating PAM imaging.

Our method focuses on the SR of single undersampled PAM images without considering scenarios wherein PAM images are acquired repeatedly on the same \textit{in vivo} subject. In future research, we will consider incorporating this additional information, as demonstrated in recent works \cite{Bhat_2021_CVPR, Dudhane_2022_CVPR}, into our framework to further improve the quality and accuracy of PAM images' SR.


\end{document}